\documentstyle[aps,pre,epsfig]{revtex}

\def\Pm{{\mathcal P}}
\def\Pb{{\mathbf P}}

\def\bea{\begin{eqnarray}}
\def\eea{\end{eqnarray}}
\def\beq{\begin{equation}}
\def\eeq{\end{equation}}
\def\Bt{{\tilde B}}
\def\mki{\langle k_i \rangle}

\begin{document}
\title{PREFERENTIAL GROWTH: EXACT SOLUTION OF THE TIME DEPENDENT DISTRIBUTIONS}
\author{L. Kullmann$^1$, J. Kert\' esz$^{1,2}$}
\address{$^1$Department of Theoretical Physics, Institute of Physics, 
Technical University of Budapest, 8 Budafoki \'ut, H-1111 Budapest, Hungary}
\address{$^2$Laboratory of Computational Engineering, Helsinki
University of Technology, P.O.Box 9400, FIN-02015, Espoo, Helsinki,
Finland}
\date{\today}
\maketitle
%%%%%%%%%%%%%%%%%%%%%%%%%%%%%%%%%%%%%%%%%%%%%%%%%%%%%%%%%%%%%%%%
\begin{abstract}
We consider a preferential growth model where particles are added one
by one to the system consisting of clusters of particles. A new
particle can either form a new cluster (with probability $q$) or join
an already existing cluster with a probability proportional to the
size thereof. We calculate exactly the probability $\Pm_i(k,t)$
that the size of the $i$-th cluster at time $t$ is $k$.  We analyze
the asymptotics, the scaling properties of the size distribution and
of the mean size as well as the relation of our system to recent
network models.

\end{abstract}
%%%%%%%%%%%%%%%%%%%%%%%%%%%%%%%%%%%%%%%%%%%%%%%%%%%%%%%%%%%%%%%%
%%%%%%%%%%%%%%%%%%%%%%%%%%%%%%%%%%%%%%%%%%%%%%%%%%%%%%%%%%%%%%%%
\section{Introduction}

Nonuniform growth is inherently present in a broad class of phenomena
including the development of biological populations, communication
networks or economic systems like incomes of persons or companies
\cite{BARAB1,BARAB2,HUBER,SIMON,FAL,REDNER,BARAB3}.  In
many cases it is obvious to assume that in a system consisting of
groups or clusters of units the attachment of a new entity to one of
the groups depends on the already achieved strength or size of that
particular group. Simon ~\cite{SIMON} analysed a simple model of this
kind where the growth probability was proportional to the cluster size
and he gave exact results for the time dependent size
distribution. Referring to the examples of words in a book or personal
incomes Simon derived a power law distribution of cluster sizes.
Recently, in the search for an explanation of the widely observed
scale invariance of large networks like the WWW
\cite{BARAB1,BARAB2,HUBER}, the Internet or power networks \cite{FAL},
scientific citation \cite{REDNER}
the idea of preferential growth has been applied to evolving graphs
\cite{BARAB3}.  It turned out that such graphs behave remarkably: They have
``small world'' properties \cite{KOCHEN} and the distribution of the
strength of vertices (number of edges from or to a vertex) is scale
free, provided that the probability of linking a vertex with a new one
is proportional to its strength \cite{KRAP} This class of models
represent a new mechanism for ``self-organized
criticality''\cite{SOC}. The idea of preferential growth seems to be
essential in economic systems too where clustering of companies, e.g.,
according to their market seem to follow such a pattern \cite{KIYO}.

These models have been treated by different tools including
simulations, continuum or mean field theories \cite{BARAB4}
and exact calculations \cite{SIMON,DOROG} by which information has
been accumulated about the asymptotic behavior and the time dependence
of the global distribution functions. However, much less attention has
been paid to the full time-dependent solution of the problem. The aim
of the present work is to give such a solution of a particular model.

The paper is organized as follows. In Section II we define the model
and the quantities of interest as well as we present the basic master
equation. In Section III the main steps of the full time dependent
analytic solution is given and the consequences for the steady state
and the integrated distributions are drawn. Section contains the
analysis about the asymptotics and scaling. In Section V we present a
discussion of our results. The paper terminates with two appendices
containing some details of the calculations.
%%%%%%%%%%%%%%%%%%%%%%%%%%%%%%%%%%%%%%%%%%%%%%%%%%%%%%%%%%%%%%%%
%%%%%%%%%%%%%%%%%%%%%%%%%%%%%%%%%%%%%%%%%%%%%%%%%%%%%%%%%%%%%%%%
\section{Model}
\begin{figure}[ht]
\centerline{
\epsfig{file=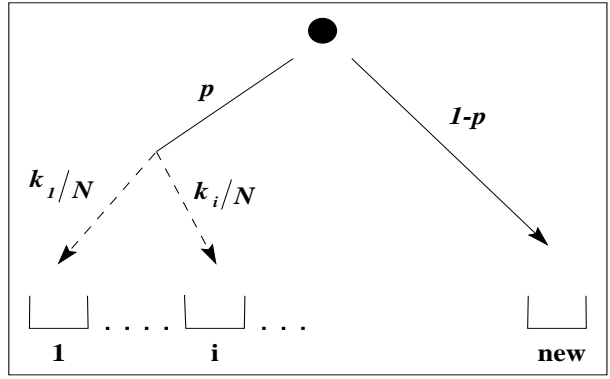,width=8truecm,height=5truecm}}
\caption{Demonstration of the model. The black point on the top denotes the 
new incoming element, the boxes on the bottom are the groups.}
\label{fig_explain}
\end{figure}

We model a growing system which consists of groups of different sizes.
At the beginning ($t=1$) we have one group with one element in it.
At each time step we add a new element to the system. With probability $p$
it will belong to one of the existing groups. The probability that it joins
the $i$-th group is proportional to the size of the group ($k_i/N$), 
see Fig~\ref{fig_explain}.  
(The number of elements is equal to the time, $N=t$, because the system size 
is rising by one in each time step.)
With probability $q=1-p$ the new element will belong to a new group.
The process can be described by the following master equation:
\bea
\label{master}
\Pm_i(k,t)
&=& p{(k-1) \over t-1}\, \Pm_i(k-1,t-1)
+ p\left(1-{k \over t-1}\right)\, \Pm_i(k,t-1) + \nonumber\\
&&+\ (1-p)\, \Pm_i(k,t-1)
+ (1-p)\, \Pi_{i-1}(t-1)
\, \delta_{k,1}(1-\delta_{i,1}),
\eea
where $\Pm_i(k,t)$ is the probability that at time $t$ there are $k$ elements
in the group $i$, and $\Pi_i(t)$ is the probability that at time $t$ there are
$i$ groups in the system:
\bea
\label{boxprob}
\Pi_i(t) = {t-1 \choose i-1}\ p^{t-1-(i-1)}\ (1-p)^{i-1}.
\eea
In the following we introduce some important quantities
and their definitions.

Given the size distribution of the individual groups, $\Pm_i(k,t)$,
the size distribution of the total system can be calculated as their
average:
\beq
\label{def_P_k,t}
\Pb(k,t) = {1 \over t}\, \sum_{i=1}^t\ \Pm_i(k,t).
\eeq
In the long time limit this quantity approximates to a stationary value:
$\Pb(k) = \lim_{t \to \infty} \Pb(k,t)$.

The mean of the $i$-th group size:
\bea
\label{def_mean}
\mki\, (t) = \sum_{k=1}^{t-i+1}\ k\ \Pm_i(k,t).
\eea
The reason that the upper limit of the above sum is not infinity is that
$\Pm_i(k,t) = 0$ if $k>t-i+1$.
%%%%%%%%%%%%%%%%%%%%%%%%%%%%%%%%%%%%%%%%%%%%%%%%%%%%%%%%%%%%%%%%%%%%%%
%%%%%%%%%%%%%%%%%%%%%%%%%%%%%%%%%%%%%%%%%%%%%%%%%%%%%%%%%%%%%%%%%%%%%%
\section{Analytic calculations}
%%%%%%%%%%%%%%%%%%%%%%%%%%%%%%%%%%%%%%%%%%%%%%%%%%%%%%%%%%%%%%%%%%%%%%
\subsection{Asymptotic distribution of group size}
\label{sec_P_k,inf}
In the first step we calculate the group size distribution in the
asymptotic case, $\Pb(k)$. 

The exact analytic formula for $\Pb(k)$ was already calculated in
\cite{SIMON,DOROG}, we present it here to see the dependence of the
exponent on the parameter $p$.

If we sum up Eq.~(\ref{master}) for $i=1\dots t$, we get:
\beq
\label{recurs_P}
t\ \Pb(k,t) = (t-1-pk)\, \Pb(k,t-1) + p(k-1)\, \Pb(k-1,t-1) + 
(1-p)\delta_{k,1}\ ,
\eeq
since:
\bea
\sum_{i=1}^{t} \Pi_{i-1}(t-1) (1-\delta_{i,1}) &=& 1, \nonumber \\
\sum_{i=1}^{t} \Pm_i(k,t-1) = \sum_{i=1}^{t-1} \Pm_i(k,t-1) &=& 
(t-1) \Pb(k,t-1).
\nonumber 
\eea
The stationary behavior of $\Pb(k,t)$, mentioned in the previous section,
can be checked from Eq. (\ref{recurs_P}).
Replacing the stationary quantity $\Pb(k)$ into Eq.~(\ref{recurs_P}) one gets:
\beq
\Pb(k) = -pk\Pb(k) + p(k-1)\Pb(k-1) + (1-p) \delta_{k,1},
\eeq
which can be solved for $\Pb(k)$:
\beq
\label{P_k,inf}
\Pb(k) = {\Gamma(k)\Gamma \left(2+{1 \over p} \right) 
\over \Gamma \left(k+1+{1\over p} \right) }\,
{1-p \over 1+p}\quad \mathop{\sim}_{k \to \infty} k^{-1-1/p}.
\eeq
\begin{figure}[ht]
\label{fig_P_k,inf}
\centerline{
\epsfig{file=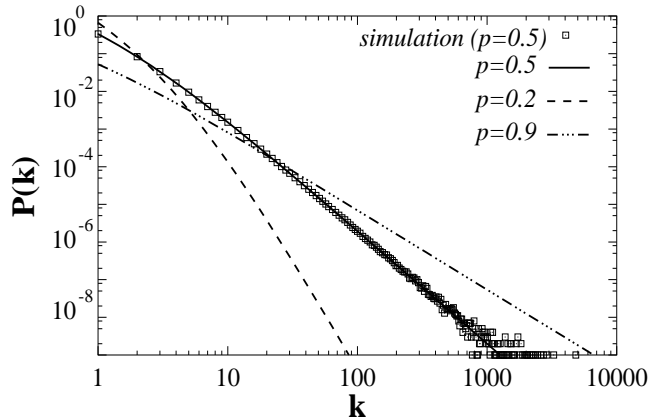,width=8.6truecm,height=5.5truecm}}
\caption{Group size distribution in the asymptotic limit, for
different $p$ values.}
\end{figure}
%%%%%%%%%%%%%%%%%%%%%%%%%%%%%%%%%%%%%%%%%%%%%%%%%%%%%%%%%%%%%%%%
\subsection{Analytic solution for the individual group size distribution}
\label{sec_Pm_i,k,t}
In the model the first group has an accentuated role since it always has 
at least one element because of the initial conditions.
Therefore the master equation (\ref{master}) for the first group ($i=1$)
has the following simpler form:
\beq
\label{i1_master}
\Pm_1(k,t) = \Pm_1(k,t-1)\ -\ {p \over t-1}\ k\ \Pm_1(k,t-1)\ +\
{p\over t-1}\ (k-1)\ \Pm_1(k-1,t-1).
\eeq
For $k=1$ in the above equation on the r.h.s the last term vanishes so the
probability $\Pm_1(1,t)$ can be calculated easily:
\beq
\Pm_1(1,t) = {\Gamma(t-p) \over \Gamma(t) \Gamma(1-p)}.
\eeq
%%%
For $k>1$ one can prove (see Appendix \ref{sec_sum}) that the following 
equality holds:
\beq
\label{k>1_i1_biz}
\sum_{k=1}^l\ (-1)^{k-1}\ {l-1 \choose k-1}\ \Pm_1(k,t) = 
{\Gamma(t-lp) \over \Gamma(t) \Gamma(1-lp)}.
\eeq

The analytic form of $\Pm_1(k,t)$ can be received from Eq.~(\ref{k>1_i1_biz})
by multiplying both sides with $(-1)^{l-1}{k-1 \choose l-1}$ and summing
up for $l=1 \dots k$,
\beq
\label{Pm_1,k}
\Pm_1(k,t) = \sum_{l=1}^k\ (-1)^{l-1}\ {k-1 \choose l-1}\
{\Gamma(t-lp) \over \Gamma(t) \Gamma(1-lp)}.
\eeq
%%%
In the case of $i>1$ we have to look at the hole Master equation
(\ref{master}). 
In this case the equality (\ref{k>1_i1_biz}) doesn't
hold because of the last factor in (\ref{master}).
Our assumption is that the probability $\Pm_i(k,t)$ will have a modified form:
\bea
\label{Pm_i,k,t}
\Pm_i(k,t)\ =\
\overbrace{
\sum_{l=1}^k\ (-1)^{l-1}\ {k-1 \choose l-1}\ 
{\Gamma(t-lp) \over \Gamma(t)\Gamma(1-lp)}
}^{\displaystyle \Pm_1(k,t)}
\left[\ \sum_{b=i}^{t}\ {\Gamma(b)\Gamma(1-lp) \over \Gamma(b-lp)}\
{b-2 \choose i-2}\ p^{b-i}\ (1-p)^{i-1}\ \right].
\eea
The validity of the above form can be checked by replacing it back
in Eq. (\ref{master}), see Appendix \ref{sec_i>1}.
%%%%%%%%%%%%%%%%%%%%%%%%%%%%%%%%%%%%%%%%%%%%%%%%%%%%%%%%%%%%%%%%
\subsection{Mean value of group sizes}
\label{sec_mean-k,t}
Replacing the analytic formula (\ref{Pm_i,k,t}) into (\ref{def_mean}) 
one gets:
\bea
\mki\, (t) = \sum_{k=1}^{t-i+1} k
\sum_{l=1}^k\ (-1)^{l-1} {k-1 \choose l-1} 
{\Gamma(t-lp) \over \Gamma(t)\Gamma(1-lp)}
\left[ \sum_{b=i}^{t} {\Gamma(b)\Gamma(1-lp) \over \Gamma(b-lp)}
{b-2 \choose i-2} p^{b-i} (1-p)^{i-1} \right],
\eea
The two sums can be transposed 
$\big( \sum_{k=1}^{t-i+1} \sum_{l=1}^k = 
\sum_{l=1}^{t-i+1} \sum_{k=l}^{t-i+1}\big)$,
and
$$
\sum_{k=l}^{t-i+1} k {k-1 \choose l-1} = l {t-i+2 \choose l+1}
$$
so the mean value will have the following form:
\beq
\label{mean-k,t}
\mki\, (t) =
\sum_{l=1}^{t-i+1} (-1)^{l-1} l {t-i+2 \choose l+1}
{\Gamma(t-lp) \over \Gamma(t)\Gamma(1-lp)}
\sum_{b=i}^t {\Gamma(b)\Gamma(1-lp) \over \Gamma(b-lp)}
{b-2 \choose i-2} p^{b-i} (1-p)^{i-1}.
\eeq
%%%%%%%%%%%%%%%%%%%%%%%%%%%%%%%%%%%%%%%%%%%%%%%%%%%%%%%%%%%%%%%%
\subsection{Time dependent solution for the group size distribution
$\Pb(k,t)$}
\label{sec_P_k,t}
In Sec. \ref{sec_P_k,inf} we calculated the stationary group size
distribution directly from the master equation. Now we are interested
in its dynamic. In order to compute that, we start from the definition 
(\ref{def_P_k,t}) of $\Pb(k,t)$, and replace the solution we got for
$\Pm_i(k,t)$ in the previous sections (\ref{Pm_i,k,t}).
\bea
\label{P_k,t__}
\Pb(k,t) &=& {1 \over t} \sum_{l=1}^k (-1)^{l-1} {k-1 \choose l-1}
{\Gamma(t-lp) \over \Gamma(t)\Gamma(1-lp)}
\left[1 +
(1-p) \sum_{i=2}^t\ \sum_{b=i}^t {\Gamma(b)\Gamma(1-lp) \over \Gamma(b-lp)}
{b-2 \choose i-2}\ p^{b-i}\ (1-p)^{i-2} \right]
\eea
Transposing the two sums: 
$\sum_{i=2}^t \sum_{b=i}^t = \sum_{b=2}^t \sum_{i=2}^b$, and taking into
account that:
$$
\sum_{b=2}^t {\Gamma(b)\Gamma(1-lp) \over \Gamma(b-lp)}\ =\ 
{\Gamma(1-lp) \over (1+lp)}\
{\Gamma(t+1) \over \Gamma(t-lp)}\ -\ {1 \over 1+lp},
$$
one finally arrives at the time dependent distribution:
\bea
\label{P_k,t}
\Pb(k,t)\ =\ 
\underbrace{
\sum_{l=1}^k\ (-1)^{l-1}\ {k-1 \choose l-1}\ \left[ {1-p \over 1+lp}\ \right.
}_{\displaystyle \Pb(k,\infty)}
+\ {p+lp \over 1+lp}\
{\Gamma(t-lp) \over \Gamma(t+1)\Gamma(1-lp)}
\Bigg]
\eea
In the long time limit we will get back our result (\ref{P_k,inf})
since the second term in $\big[\dots \big]$ decays for large $t$ values
with $t^{-1-lp}$, and the sum transforms into:
\bea
\Pb(k,\infty)\ =\ {1-p \over 1+p}\ 
{\Gamma(k)\Gamma(2+1/p) \over \Gamma(k+1+1/p)} \nonumber.
\eea
%%%%%%%%%%%%%%%%%%%%%%%%%%%%%%%%%%%%%%%%%%%%%%%%%%%%%%%%%%%%%%%%
%%%%%%%%%%%%%%%%%%%%%%%%%%%%%%%%%%%%%%%%%%%%%%%%%%%%%%%%%%%%%%%%
\section{Asymptotic cases}
We study the $t \to \infty$ limes of $\Pm_i(k,t)$ and
$\mki (t)$.
In the analytic formula for $\Pm_i(k,t)$, see. Eq. (\ref{Pm_i,k,t}), there
are two components, $A(l,p,t)$ and $B(i,l,p,t)$,
that depend on the time.
\bea
\label{beta_Pm_i,k,t}
\Pm_i(k,t) =
(1-p)^{i-1}
\sum_{l=1}^k (-1)^{l-1} {k-1 \choose l-1}
\underbrace{{\Gamma(t-lp) \over \Gamma(t)\Gamma(1-lp)}
}_{\displaystyle A(l,p,t)}
\underbrace{
\left[\ 
\sum_{b=i}^{t}\ {\Gamma(b)\Gamma(1-lp) \over \Gamma(b-lp)} {b-2 \choose i-2}
p^{b-i} \right]
}_{\displaystyle B(i,l,p,t)}
\eea
The limes of the first term, $A(l,p,t)$, can be easily calculated
\bea
\lim_{t \to \infty} A(l,p,t) = {1 \over \Gamma(1-lp)} t^{-lp}
\nonumber
\eea
The second term in the long time limit $t \gg i,l$ will converge to a
hypergeometric sum:
\bea
\lim_{t \to \infty} B(i,l,p,t) = \Bt(i,l,p) = 
{\Gamma(i) \Gamma(1-lp) \over \Gamma(i-lp)}
\ _2F_1(i,i-1;i-lp;p).
\nonumber
\eea

For large time values the only time dependent term in (\ref{beta_Pm_i,k,t})
will be $t^{-lp}$ which in case of large $t$ is a fast decaying function of 
$l$. So in the case of $t \gg k$ we can assume that only the first term
of the sum gives non-negligible component for
$\Pm_i(k,t)$.
\bea
\label{Pm_i,k,inf}
\lim_{t \to \infty} \Pm_i(k,t) =
t^{-p}\ (1-p)^{i-1}
{\Gamma(i) \over \Gamma(i-p)}\ {_2}F_1 (i,i-1;i-p;p)
 + {\mathcal O}(t^{-2p})
\eea
For large $i$ values the above formula simplifies further, because
in that case $\lim_{i \to \infty} {_2}F_1 (i,i-1;i-p;p) \sim (1-p)^{1-i}$,
and $\lim_{i \to \infty} {\Gamma(i) \over \Gamma(i-p)} = i^{p}$:
\beq
\label{Pm_inf,k,inf}
\lim_{t,i \to \infty} \Pm_i(k,t) = \left( {i \over t} \right)^p
\eeq

\begin{figure}[ht]
\centerline{
\epsfig{file=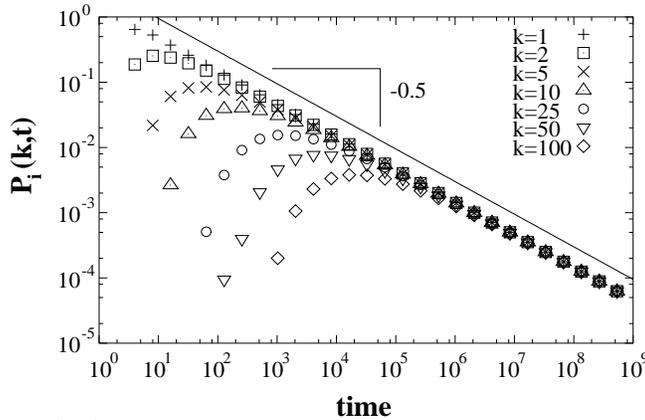,width=8.6truecm,height=5.5truecm}}
\caption{Asymptotic behavior of $\Pm_i(k,t)$.
We chose the parameters for $p=0.5$ and $i=2$. The figure
demonstrates that in the long time limit the probabilities for
different $k$ values converge to $t^{-p}$.}
\label{fig_conv_Pm_i,k,t}
\end{figure}

To study the asymptotic behavior of $\mki(t)$
we start from the fact, that for small $k$ values, $k \ll t$, the individual
group size distribution, $\Pm_i(k,t)$, can be described by the
first term of the sum, see Eq. (\ref{Pm_i,k,inf}), and for larger
values $k \gtrsim t$ it has a fast decay, Fig. \ref{fig_pm-k,inf}.
A cut-off parameter, $k^*$, can be defined and we can assume that
(\ref{def_mean}) transforms into
\beq
\label{mean_inf}
\mki (t) \approx \sum_{k=1}^{k^*} k\ \Pm_i(k,t) =
\Pm_i(1,t) {k^* (k^* +1) \over 2}.
\eeq
\begin{figure}[ht]
\centerline{
\epsfig{file=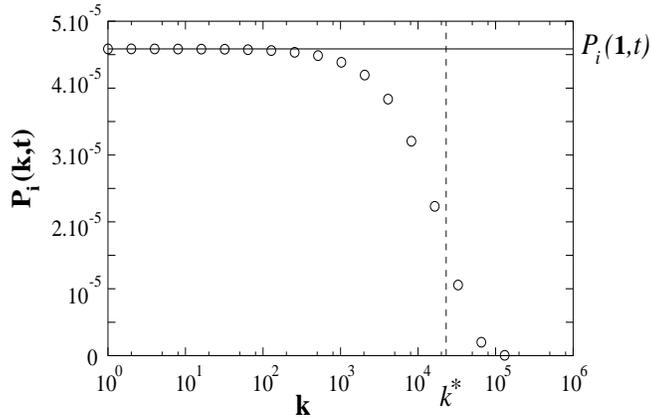,width=8.6truecm,height=5.5truecm}}
\caption{Distribution of individual group size in the long time
limit ($t = 10^9$) as a function of the group size}
\label{fig_pm-k,inf}
\end{figure}
The definition of $k^*$ can be done in many ways. We defined $k^*$
as the inflection point of $\Pm_i(k,t)$, hence:
\beq
\label{def_k_star}
k^* = t^p\ {\Bt(i,3,p) \over \Bt(i,4,p)}\ {\Gamma(1-4p) \over \Gamma(1-3p)} 
+ 2 + {\mathcal O}(t^{-p})= t^p\ {\Gamma(i-4p) \over \Gamma(i-3p)}\
{{_2}F_1(i,i-1;i-3p;p) \over {_2}F_1(i,i-1;i-4p;p)} + 2 + {\mathcal O}(t^{-p}).
\eeq
Replacing $k^*$ into (\ref{mean_inf})
\beq
\mki (t) \simeq t^p (1-p)^{i-1} 
{\Gamma(i) \over \Gamma(i-p)}\ {_2}F_1 (i,i-1;i-p;p)
\left[ {\Gamma(i-4p) \over \Gamma(i-3p)}\
{{_2}F_1(i,i-1;i-3p;p) \over {_2}F_1(i,i-1;i-4p;p)} \right]^2\ .
\eeq
For large $i$ values the above formula gets a simpler form, because
in this case $\lim_{i \to \infty} {\Gamma(i) \over \Gamma(i-p)} = i^{p}$,
$\lim_{i \to \infty} {\Gamma(i-4p) \over \Gamma(i-3p)} = i^{-p}$,
$\lim_{i \to \infty}{_2}F_1(i,i-1;i-3p;p) 
= \lim_{i \to \infty}{_2}F_1(i,i-1;i-4p;p)$,
$\lim_{i \to \infty} {_2}F_1 (i,i-1;i-p;p) \sim (1-p)^{1-i}$.
\beq
\label{mki_inf}
\mki (t)\ \mathop{\approx}_{i \to \infty}\ \left( {t \over i} \right)^p
\eeq
%%%%%%%%%%%%%%%%%%%%%%%%%%%%%%%%%%%%%%%%%%%%%%%%%%%%%%%%%%%%%%%%
%%%%%%%%%%%%%%%%%%%%%%%%%%%%%%%%%%%%%%%%%%%%%%%%%%%%%%%%%%%%%%%%
\section{Discussion}
In this paper we presented a simple preferential growth model
consisting of a system of clusters with different sizes.  We gave
exact solutions for the main characteristic quantities as the
distribution, $\Pm_i(k,t)$, and the mean value, $\mki (t)$, of the
individual group size as well as for the distribution of the
average group size, $\Pb(k,t)$.

The question rises why are such time dependent quantities of interest
since most of the asymptotic scaling behavior can be obtained with
much less labor. In fact the growth models and network usually provide
only a background for some dynamic process -- an aspect which has not
yet paid enough attention to. If there is a strong separation of time
scales, i.e., the growth is much smaller than the process itself then 
it is satisfactory to concentrate on the asymptotics only. This is
probably the case with the Internet or the WWW. However, in some cases 
such a separation of scales could be approximate only or even missing
and then the importance of the full time dependence becomes apparent.
We expect that in certain economic processes this will be the case.

An important aspect in the asymptotic scaling is
universality. Similarly to other preferential growth models, our
system exhibits nonuniversal parameter dependent scaling: the
exponents depend on the parameter $q$ (the probability of creating a
new group). It is worth mentioning that the examples quoted in the
introduction also show a wide variety of scaling exponents. Further
interesting study would be to analyse a model where this parameter $q$
depends on the time of the growth.

The presented system is not a network, the different groups are not
linked to each other. However, for a specific value of the parameter,
$p=0.5$, it can be interpreted as a kind of mean field network
model. The clusters then denote the different nodes, and the particles
are the links.  The value $p=0.5$ means that in average in every
second time step one new group and two elements are created (in the
odd time steps the new element joins to an old group and in even time
steps it will create a new group.) The new group is the new node while
the two new elements are the two ends of the new link, one is pointing
to the old node, the other is to the new one.  This case corresponds
to the Barabasi's network model with parameter $m=1$ which means that
the new node connects to {\it one} old sites.  For this particular
parameter choice our results agree with them got for the Barabasi's
network model: $\Pb(k) \sim k^{-3}$, see Eq. (\ref{P_k,inf}), and
$\mki (t) \sim \sqrt{t/i}$, see Eq. (\ref{mki_inf}).
\vskip 0.5truecm

{\bf Acknowledgement:} This research was supported by OTKA T029985.

%%%%%%%%%%%%%%%%%%%%%%%%%%%%%%%%%%%%%%%%%%%%%%%%%%%%%%%%%%%%%%%%
%%%%%%%%%%%%%%%%%%%%%%%%%%%%%%%%%%%%%%%%%%%%%%%%%%%%%%%%%%%%%%%%

%%%%%%%%%%%%%%%%%%%%%%%%%%%%%%%%%%%%%%%%%%%%%%%%%%%%%%%%%%%%%%%%
%%%%%%%%%%%%%%%%%%%%%%%%%%%%%%%%%%%%%%%%%%%%%%%%%%%%%%%%%%%%%%%%
\appendix
\section{}
\label{sec_sum}
We prove the assumption (\ref{k>1_i1_biz}).

If one multiplies (\ref{i1_master}) by $(-1)^{k-1}{l-1 \choose k-1}$ and sums
it up for $k=1 \dots l$ one gets:
\bea
\label{k>1_i1_rekurz}
&& \sum_{k=1}^l\ (-1)^{k-1}\ {l-1 \choose k-1}\ \Pm_1(k,t) =
\sum_{k=1}^l\ (-1)^{k-1}\ {l-1 \choose k-1}\ \Pm_1(k,t-1)\ -\ 
\nonumber \\
&& -\ {p \over t-1}
\Bigg[
\underbrace{\sum_{k=1}^l (-1)^{k-1} {l-1 \choose k-1} k\
\Pm_1(k,t-1)}_{\textrm x}\
-\
\underbrace{\sum_{k=1}^l (-1)^{k-1} {l-1 \choose k-1} (k-1)
\Pm_1(k-1,t-1)}_{\textrm y}
\Bigg],
\eea
where in the first term ($x$) we detach the last term of the sum:
\beq
x = \sum_{k=1}^{l-1} (-1)^{k-1}\ k {l-1 \choose k-1}  \Pm_1(k,t-1) +
(-1)^{l-1}\ l \Pm_1(l,t-1).
\eeq
Taking into account that ${l-1 \choose k} = {l-k \over k}{l-1 \choose k-1}$
the second term ($y$) can be rewritten as:
\bea
y = \sum_{k=2}^l (-1)^{k-1} {l-1 \choose k-1} (k-1) \Pm_1(k-1,t-1) =
 -\ \sum_{k=1}^{l-1} (-1)^{k-1}\  (l-k)
{l-1 \choose k-1} \Pm_1(k,t-1).
\eea
\beq
\label{x_min_y}
x-y = l\ \sum_{k=1}^l (-1)^{k-1} {l-1 \choose k-1} \Pm_1(k,t-1).
\eeq
Replacing the difference, $x-y$, back to Eq.~(\ref{k>1_i1_rekurz}) one
gets the time evolution of the sum:
\bea
\sum_{k=1}^l\ (-1)^{k-1} {l-1 \choose k-1} \Pm_1(k,t)\ =
{t-1-lp \over t-1}
\sum_{k=1}^l\ (-1)^{k-1} {l-1 \choose k-1} \Pm_1(k,t-1),
\eea
which leads us back to our assumption (\ref{k>1_i1_biz}).
%%%%%%%%%%%%%%%%%%%%%%%%%%%%%%%%%%%%%%%%%%%%%%%%%%%%%%%%%%%%%%%%
\section{}
\label{sec_i>1}
We prove the formula (\ref{Pm_i,k,t}) for $\Pm_i(k,t)$ in the case of
$i>1$, by replacing it into Eq.~(\ref{master}).

The l.h.s of the equation after detaching the last term ($b=t$) of the sum:
\bea
\label{1term}
\Pm_i(k,t) = \sum_{l=1}^k (-1)^{l-1} {k-1 \choose l-1}\
{t-2 \choose i-2} p^{t-i} +
\sum_{l=1}^k (-1)^{l-1} {k-1 \choose l-1}
{\Gamma(t-lp) \over \Gamma(t)\Gamma(1-lp)}
\sum_{b=i}^{t-1} {\Gamma(b)\Gamma(1-lp) \over \Gamma(b-lp)}
{b-2 \choose i-2} p^{b-i}\ .
\eea
The first term of the r.h.s:
\bea
\label{2term}
{t-1-kp \over t-1}\ \Pm_i(k,t-1) &=&
(-1)^{k-1} {\Gamma(t-kp) \over \Gamma(t)}
\sum_{b=i}^{t-1}\ {\Gamma(b) \over \Gamma(b-kp)}
{b-2 \choose i-2} p^{b-i}\ + \nonumber \\
&+& {t-1-kp \over t-1}\ 
\sum_{l=1}^{k-1}\ (-1)^{l-1}\ {k-1 \choose l-1}
{\Gamma(t-1-lp) \over \Gamma(t-1)}
\sum_{b=i}^{t-1}\ {\Gamma(b) \over \Gamma(b-lp)}
{b-2 \choose i-2} p^{b-i},
\eea
Taking into account that
$(k-1)\ {k-2 \choose l-1} = (k-l)\ {k-1 \choose l-1}$,
the second term will be:
\bea
\label{3term}
{(k-1)p \over t-1} \Pm_i(k-1,t-1) =
{p \over t-1}
\sum_{l=1}^{k-1} (-1)^{l-1} (k-l) {k-1 \choose l-1}
{\Gamma(t-1-lp) \over \Gamma(t-1)}
\sum_{b=i}^{t-1} {\Gamma(b) \over \Gamma(b-lp)}
{b-2 \choose i-2} p^{b-i}.
\eea

The sum of (\ref{2term}) and (\ref{3term}) will be:
\bea
\label{23term}
&&{t-1-kp \over t-1}\ \Pm_i(k,t-1) + {(k-1)p \over t-1}\ \Pm_i(k-1,t-1) = 
\nonumber\\
&&=\ (-1)^{k-1} {\Gamma(t-kp) \over \Gamma(t)}
\sum_{b=i}^{t-1}\ {\Gamma(b) \over \Gamma(b-kp)}
{b-2 \choose i-2} p^{b-i}\ +
\sum_{l=1}^{k-1}\ (-1)^{l-1}\ {k-1 \choose l-1}\ 
{\Gamma(t-lp) \over \Gamma(t)}\
\sum_{b=i}^{t-1}\ {\Gamma(b) \over \Gamma(b-lp)}\
{(b-2) \over (i-2)}\ p^{b-i} = \nonumber \\
&&=\ \sum_{l=1}^k\ (-1)^{l-1}\ {k-1 \choose l-1}\ 
{\Gamma(t-lp) \over \Gamma(t)}\
\sum_{b=i}^{t-1}\ {\Gamma(b) \over \Gamma(b-lp)}\
{(b-2) \over (i-2)}\ p^{b-i},
\eea
which will be equal to the second term of (\ref{1term}). 
Simplifying with this
term the remaining equation:
\beq
{t-2 \choose i-2}p^{t-i}\ 
\sum_{l=1}^k\ (-1)^{l-1}\ {k-1 \choose l-1}\ =
{t-2 \choose i-2}\ p^{t-i}\ \delta_{k,1}.
\eeq
Which is true, because the sum equals with $\delta_{k,1}$.

\end{document}